# THIRTY YEARS OF TEFLIN JOURNAL: A BIBLIOMETRIC PORTRAIT THROUGH THE LENS OF MICROSOFT ACADEMIC

**Abdul Syahid[a], Nur Mukminatien[b]**
([a]abdul.syahid@iain-palangkaraya.ac.id, [b]nur.mukminatien.fs@um.ac.id)

*[a]Institut Agama Islam Negeri Palangka Raya*
*Jalan G Obos, Kompleks Islamic Center, Palangka Raya 73112, Indonesia,*

*[b]Universitas Negeri Malang*
*Jalan Semarang No.5, Malang 65145, Indonesia*

**Abstract:** To examine bodies of literature from the levels of topics, regions, nations, and journals, a lot of bibliometric studies have been conducted in many fields. However, such studies are a rare undertaking in the field of English language teaching, especially at a journal level. To celebrate the TEFLIN Journal's 30th anniversary, this study exhibits a bibliometric portrait of its publication, indexation, and citation from 1990 to 2019. Two pieces of free software, Publish or Perish and VOSviewer were adopted to conduct the descriptive and network analyses of bibliographic data from Microsoft Academic. In terms of 19 out of 27 metrics in Publish or Perish software, the journal's publication and citation metrics have risen during its lifetime. The bibliographic network identifies the most productive authors, institutions, and countries along with the co-authorship pattern, type of top-cited articles, and top-used keywords. The articles relatedness is also sighted in terms of the citation frequency and number of shared references. Even though the analyses were complicated by some missing articles and improper indexation, this study could still take a full-length bibliometric portrait of the journal during its 30-year journey between the commitment to competence and the quest for higher impact.

The TEFLIN Journal (TJ) was founded in 1990. It is published by the Association for the Teaching of English as a Foreign Language in Indonesia (TEFLIN) through its TEFLIN Publication Division, based in Universitas Negeri Malang, Indonesia. As stated in the TEFLIN Journal's website (http://journal.teflin.org), TJ has consistently focused on publishing and disseminating peer reviewed conceptual and research-based articles within the fields of teaching English as a second or foreign language, English language teaching and learning, English language teachers' training and education, and English language and literary studies.

Having been indexed in various reputable databases, TJ boasts the national, regional, and global recognitions of its contribution. Nationally, since 2016 TJ has been rated "A" or Sinta-1 in the Science and Technology Index (SINTA), the only Indonesian state-managed abstract and citation index (National Agency for Research and Innovation, 2021; TEFLIN Journal, 2020). Regionally, TJ's contents from 2015 to present are covered by ASEAN Citation Index/ACI (https://asean-cites.org/) , a scientific database for scholarly journals published by 10 member states of the Association of South East Asian Nations/ ASEAN (ACI, 2020). Globally, since 2010 TJ has been included in Directory of Open Access Journals/DOAJ (https://doaj.org/), an indexing service for worldwide high-quality peer-reviewed journals providing open access contents (DOAJ, 2020). TJ is one of first eight Indonesian journals and of first 18 Southeast Asian ones added in DOAJ (Syahid, 2020). The Education Resources Information Center (ERIC), an American database of educational research, also acknowledges TJ's scientific contribution to the big enterprise of education by indexing all of its publication (ERIC, 2020). Publishing not only research-based articles but also conceptual ones within the field, in 2018 TJ could reach a significant milestone as one of the Indonesian journals indexed in Scopus, one of the most reputable academic search engines and bibliographic databases (ASEBDSs). As per March 2021, there are 94 Indonesian journals indexed in Scopus (National Agency for Research and Innovation, 2021; Scopus, 2021).

Rising to the editorial and managerial challenges, TJ published its 30th volume in 2019. Motivated by the 30th publication year of TJ, this study used bibliometrics as the bedrock. Bibliometric studies had been carried out even before it was first coined as "bibliométrie" in 1934 by Paul Otlet (Rousseau, 2014, p. 218). The term was then anglicized as "bibliometrics" by Pritchard (1969, p. 349) who defines it as "the application of mathematics and statistical methods to books and other media of communication". One of the earliest bibliometric work was done by Cattell (1906) performing a statistical analysis of 1,000 American scientists and their distribution among disciplines, states, and institutions.

Nowadays bibliometrics has developed from simply tabulating scholars and their scholarly work to measuring the complex scientific productivity and impacts (Lei & Liu, 2019a). Interestingly, a Ph.D. holder in Structural Linguistics (Masic, 2017), Eugene Garfield revolutionized bibliometrics by including citation-based metrics in his Science Citation Index with its well-known phrase "impact factor" in 1964 (Garfield, 2007). The online version of the citation index could also be accessed through the Web of Science (WoS), one of the subscription-based bibliographic platforms.

Since the introduction of Science Citation Index, the three pillars of bibliometrics to assess the scientific progress from the global to individual levels have been the publication, citation, and indexation of which WoS and Scopus are the top players (Nylander et al., 2020). As previously discussed, even from the very beginning, bibliometric studies have been conducted by not only bibliometricians but also scholars from various research areas, including applied linguistics, which is closely related to the focus and scope of TJ.

Despite the increasing interest in the bibliometric studies of various disciplines, few studies have been published on the bibliometric overview of the multidisciplinary field of ELT and its related disciplines such as (applied) linguistics, Second Language Acquisition (SLA), and education. Fortunately, the few studies analyzing the bibliometric indicators of disciplines to the focus and scope of TJ offer a wide range of contexts, i.e. the global, regional, country, and journal levels.

At the global level, at least two bibliometric analyses could be found. Lei and Liu (2019a) examined approximately 10 thousand articles published between 2005 and 2016 in 42 journals included in Social Science Citation Index, one of the databases managed by WoS. They found that sociocultural issues and the use of new theories from other disciplines such as the complexity theory in the computer science were among the most popular topics in the field. More recently, Zhang (2020) examined the main trends in SLA from 1997 to 2018. The data of co-citation and keywords were retrieved from nearly 8,000 articles published by 16 leading journals registered in the bibliographic dabases of WOS platform. One of the surveyed top-tier journals was ELT Journal (https://academic.oup.com/eltj). He could identify key documents, prolific authors, leading institutions and regions along with major topics between 1997 and 2018 in SLA.

At the regional level, one bibliometric analysis is worth mentioning. In the study, Barrot (2017) paid attention to the SLA research by the researchers in Southeast Asia, the region where Indonesia as the publishing country of TJ was located. The bibliometric data of 2,706 documents by the Southeast Asian authors between 1996 and 2015 were downloaded from Scopus. Based on the five bibliometric indexes in Scopus such as H-index, she found that, even though the 10 countries just contributed to less than 2% of documents and 1% of citations in the worldwide language and linguistics enterprise, the region as a whole demonstrated an increasing scientific quality and quantity. With a total of 141 documents in Scopus, Indonesia could rank fourth in terms of productivity, i.e. 0.1% of total documents, but sixth in terms of citations, i.e. 0.011%, and seventh in terms of H-index. Universitas Negeri Malang, the university publishing house of TJ, was among the selected universities in the region actively involved in the field.

Barrot (2007) also wrote a list of 20 most preferred journals among Southeast Asian scholars of language and linguistics. TJ, however, could not be seen in the list yet because of two possible reasons. Firstly, the ASEBD adopted in her study was Scopus only. Secondly, TJ was accepted for inclusion in Scopus after the study had been published two years earlier.

As previously stated bibliometrics can also be used to measure a country’s research performance. One of the few studies in the domain of language and linguistics was carried out by Lei and Liao (2017). From Social Science Citation Index of the WoS platform, they mined bibliometric data of 1,381 articles and book reviews published from 2002 to 2012 by researchers from China and its three related territories, i.e. Hong Kong, Macao, and Taiwan. The examined decade witnesses a significant annual increase in scientific productivity by the four regions but no significant difference between them in research impacts. Among the regions, Hong Kong was the leader in terms of quality, i.e. total and annual documents, and quality, i.e. impact factors and citations. How China and its related territories wholly or partially could be power in natural and social sciences including language and linguistic was driven by the rapid economic development by which research and development expenditure was greatly supported.

So far, few researchers have done a bibliometric analysis of a single journal in the domain of language and linguistics. Among the few were Lei and Liu (2019b) who examine System (https://www.journals.elsevier.com/system), a highly reputable journal in the field (Lei & Liao, 2017; Lei & Liu, 2019a; Zhang, 2020). They explored Scopus database to mine the bibliometric data of almost 1,600 articles published by System in the 1973-2017 period. They found that over the 45 years the topics increasing significantly were related to not only classic teaching learning practice such as corrective feedback but also newly developed sociocultural and technological practice. Surprisingly, not only some outdated topics such as language laboratory and cloze test but also some key topics such as language teaching and syllabus decreased significantly. Based on the bibliometric analysis, they conclude that while the central issue in *System* is about the teaching and learning of foreign and second language learning, the journal has also paid great attention to socio-cultural and socio-psychological issues in the field during the study period. By so doing, the two authors could help, for instance, the gatekeepers such as editors and reviewers to reflect on whether the journal has achieved what it originally envisaged, i.e. solving "problems of foreign language teaching and learning" through "the applications of educational technology and applied linguistics" (System, 2020).

As can be seen in the previous studies, most of the researchers extracted the data from the two key players in the business of ASEBDs, i.e., WoS and Scopus. Both of them are subscription-based platforms. The subscription however sets a limit on the "*reproducibility* (also "replicability," "reliability," and "repeatability") and *transparency"* (Gusenbauer & Haddaway, 2020, p. 184, emphasis in original) in doing bibliometric research. Extracting bibliometric data from free ASEBDs such as Microsoft Academic/MA (https://academic.microsoft.com/) could be a viable solution to no access to the two platforms. Much work on the comparison of different bibliographic databases including MA, WoS' databases, and Scopus has been carried out (Gusenbauer & Haddaway, 2020; Harzing, 2019; Hug et al., 2017; Thelwall, 2017). Moreover, Scopus could not capture the bibliometric portrait of any journal beyond its coverage years. The results suggest that MA could be excellently used for the bibliometric analysis in terms of scope, accessibility, plausibility and usability, especially because of the Application Programming Interface for accessing the data more easily.

With regard to the 30th anniversary of TJ, the milestone is indeed worth celebrating. Because "the periodical publishing industry is not as prosperous, or its future as promising" (p. 219), what a journal has done to solve the internal and external problems including plagiarisms (Xiao-Jun et al., 2012) must be well appreciated. TJ has counteracted the findings that, among others, a cessation of journal averages three years after the launch time and that monolingual journals using foreign languages suffer much higher risk of cessation than multilingual journals and those published in native languages do (Liu et al., 2018). Furthermore, money, neither technology nor science, is the top cause of death in the journal publication (Silver, 2018). More interestingly, as one of the first Indonesian open access journals included in DOAJ (2020), TJ stands still until today with no submission fee and article processing charge. On the other hand, its focus and scope do not belong to research areas of the national priority in Indonesia (Wiryawan, 2014) to which the expenditure was more greatly spent. The three decades have therefore witnessed how, as an academic venue for the theorists and practitioners in the (English) language teaching arena, TJ could successfully stand the test of time.

This study examines the 30-year contribution TJ has made on its journey, borrowing the title of a TJ's article by Lie (2007, p. 1), "between the commitment to competence and the quest for" higher impacts. It is motivated by the three-decade continued commitment to and competence in the field as shown by the journal's stakeholders. Also, few studies have been published on the bibliometric indicators of an individual journal in the big enterprise of applied linguistics (System, 2020) including ELT in the contexts of second and foreign language. Related to the bibliometric analysis, this study investigates two clusters of research questions.

The first cluster concerns the descriptive analyses of publication, indexation, and citation. The questions are as follows.

1. How was the portrait of publication and indexation of TJ in its own website and some indexing services displayed in the website of TJ?
2. How was the bibliometric description of TJ in terms of the 27 indicators during the 30-year time span?
3. How was the bibliometric description of TJ in terms of the 27 indicators in the first, second, and third decades?

The second cluster is concerned with the network analyses of co-authorship, citation, co-occurrence, and bibliographic coupling. The questions are as follows:

1. Who were the most productive authors in TJ during the 30-year time span?

2. Who were the most productive authors in TJ in the first, second, and third decades?
3. What were the most productive institutions in TJ?
4. What are the most productive countries in TJ?
5. How was the pattern of authorship in TJ?
6. What were the top-cited articles and the article types in TJ?
7. What were the top-used keywords in TJ?
8. How was the relatedness of articles in TJ viewed from the frequency they cited at each other?
9. How was the relatedness of articles in TJ viewed from the number of shared references?

The above questions are used as the viewfinders for taking a better full-length bibliometric portrait of TJ. The portrait could tell the story of what have been achieved by "enhancing and disseminating scholarly work in the field of ELT" (TEFLIN Journal, 2020) during the 30-year lifetime.

## METHOD

To answer the first question of the first cluster, on June 9, 2020, the authors went onto the journal's website (http://journal.teflin.org/) and the indexing services in the website, i.e., DOAJ, ERIC (https://eric.ed.gov/), Scopus (https://www.scopus.com/), and SINTA (http://sinta.ristekbrin.go.id/. To answer the next questions in the first cluster, on the same day this study run a free computer program, Publish or Perish/PoP (Harzing, 2007) to mine more effectively the bibliometric data. The software was used for analyzing bibliometric indicators of publication and citation extracted from the five ASEBDs comprehensively compared by Harzing (2019). For the purpose of this study, the appropriate data sources were only four sources, i.e. Crossref (https://www.crossref.org/), Google Scholar (https://scholar.google.com/), MA, and Scopus (https://www.scopus.com/). The application was run as described by Harzing (2011) including the use of Application Programming Interface key required by MA and Scopus.

The queries used for each source in the software were *Publication name* or *Full journal name* (for MA only)*: TEFLIN Journal* and *ISSN: 0215-773X* or *2356-2641*. The *Years* query was filled with *0-0* (unspecific years), *1990-2019* (publishing years), *1990-1999* (the first decade), *2000-2009* (the second decade), *2010-2019* (the third decade). All of the queries were consecutively combined in order to increase the precision, for instance *Years: 2010-2019, Publication name: TEFLIN Journal*, and *ISSN: 2356-2641* using PoP (Figure 1). The gathered bibliometric data were then compared and cleaned for duplication and error. The procedure was aimed at finding the best ASEBD in terms of coverage and its suitability for PoP and VOSviewer, namely MA.

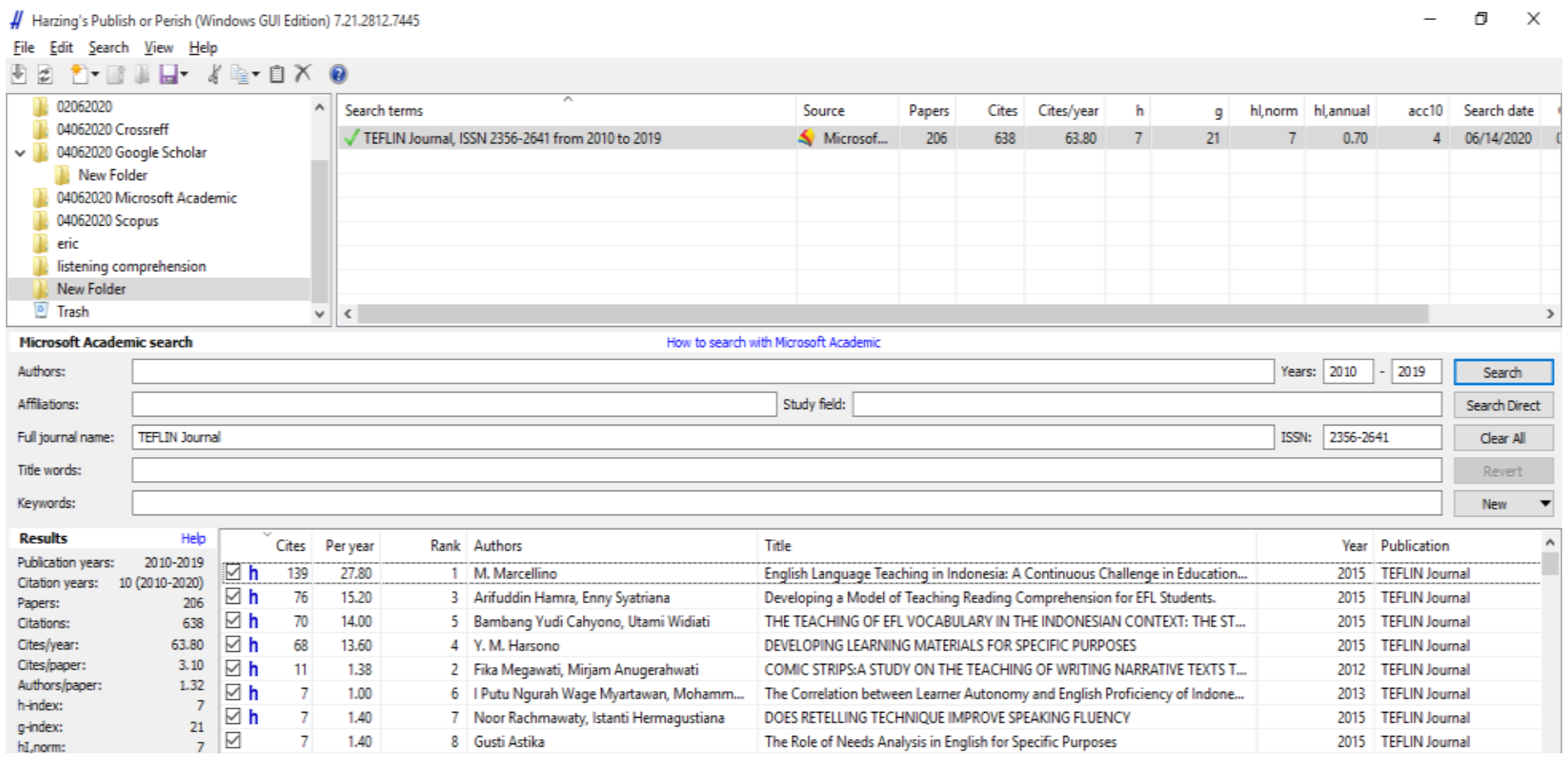

**Figure 1. Search Queries in Publish or Perish**

To answer the questions in the second cluster, VOSviewe*r* (Eck & Waltman, 2020), one of the most popular bibliometric applications (Pan et al., 2018), was used for visualizing the bibliometric similarities (connectedness) in terms of co-authorship, co-occurrence, citation, bibliographic coupling, and co-citation based on bibliographic and text data in TJ. Based on the comparison of the sources for PoP software, MA as the most reliable source was then used as the data source. As in the first software, the two queries were consecutively combined, for example *Journal: TEFLIN Journal* and *Year: 2000-2009* (Figure 2). The boxes of “Restrict to primary documents” was unchecked in order to allow MA retrieve the data not only from the journal’s website but also from other valid sources such as ERIC. As explained further in the next section, the website provided fewer articles than the journal has published. At the study period, some of the Digital Object Identifiers (DOIs) provided by the journal could not be found in the DOI system.[1] More details on the syntax of a DOI name can be found in the International DOI Foundation (2019).

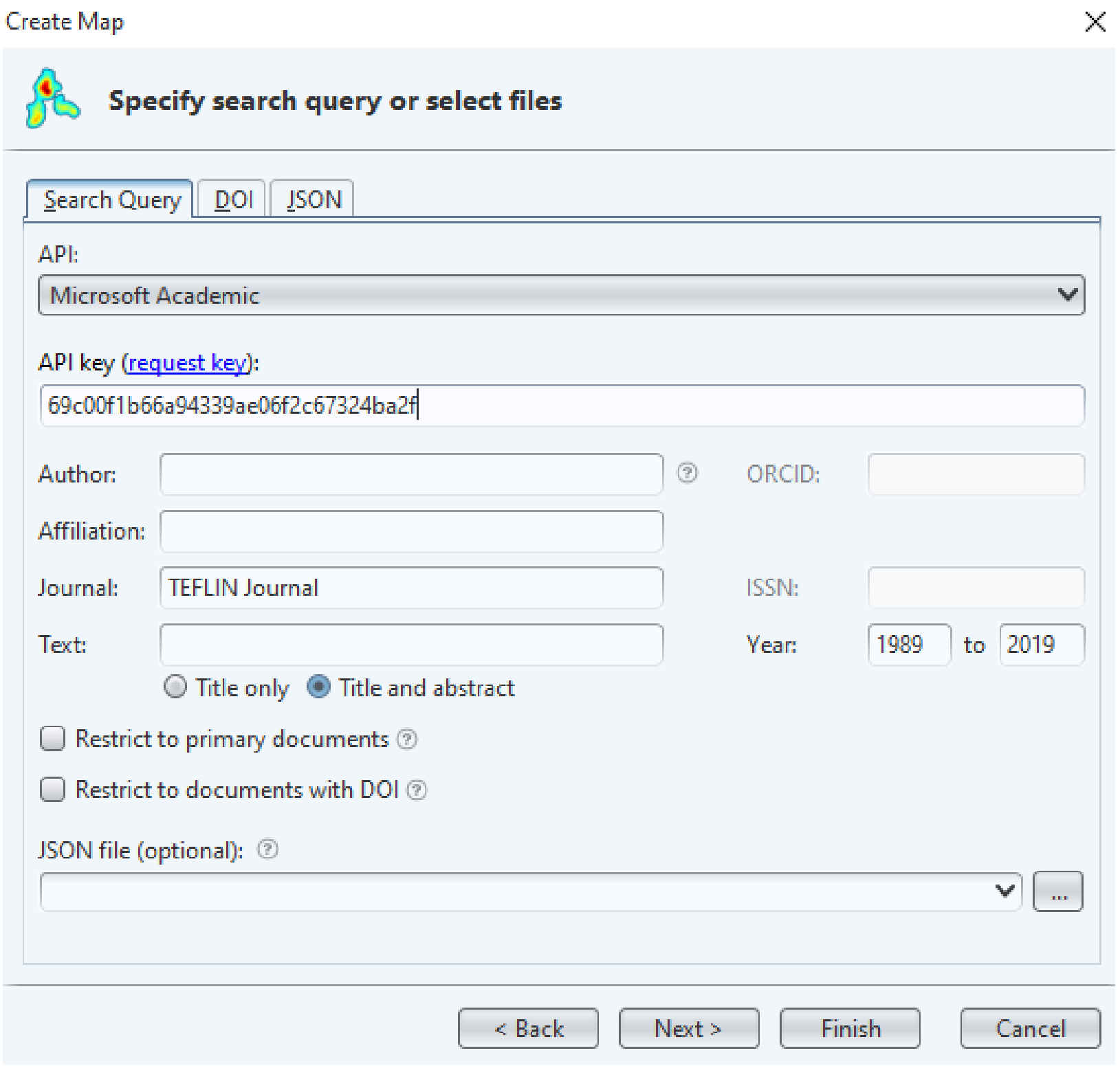

**Figure 2. Search Queries in VOSviewer**

Whereas for the co-authorship the unit of analysis was authors and organizations, for co-occurrence it was the field of study. The units of analysis for remaining analyses were documents, sources, authors, and organizations. The counting method chosen was the fractional counting by which the weight of co-authorship, citation, and co-citation was divided by the number of authors/ organizations. Each of five authors in a single article, for example, will have a co-authorship link with a weight of 1/5. Generally, the software was operated as explained by Eck and Waltman (2020).

[1] Based on e-mail correspondence with the editors of TEFLIN Journal during the preparation for this article publication, the DOI issue has now been identified and fixed.

Regarding the above procedure, this study did not involve human subjects. Thus, neither the statement of informed consent nor the approval of institutional review board was needed

## FINDINGS AND DISCUSSION

Unless otherwise specified, the findings and discussion were presented under spatial temporal considerations, namely the adopted tools and ASEBD along with the time of data retrieval. As the basis of findings and discussion, the bibliographic dataset (Syahid & Mukminatien, 2020) was provided as downloadable supplementary materials.

### Findings

#### *The Portrait of Publication, Indexation, and Citation*

The results of early exploration into six websites is presented in Table 1. However, the manually retrieved bibliographic data were not sufficient to answer the second and third research questions in the first cluster. For example, the data from ERIC contained a piece of basic information only such as authors' names. That is why the bibliometric data were retrieved from Crossref, Google Scholar, MA, and Scopus through PoP.

**Table 1. Manual Search of the TEFLIN Journal's Publication**

| Sources | Citation Index | Coverage Year | | No. of Articles |
|---|---|---|---|---|
| | | First | Last | |
| ASEAN Citation Index | No | 2015 | 2019 | 71 |
| Directory of Open Access Journals | No | 1997 | 2016 | 262 |
| ERIC | No | 1993 | 2019 | 141 |
| Scopus | Yes | 2018 | 2019 | 33 |
| Science and Technology Index | Yes | 1994 | 2015 | 216 |
| TEFLIN Journal | No | 1999 | 2019 | 290 |

The results of journal search in the four ASEBDs through PoP were refined and rechecked to find out duplication and working links. Based on the data refinement and rechecking, it could be found that the best ASEBD was MA because MA offered the best coverage and, in particular, suitability for PoP and VOSviewer.

Related to the second question of descriptive analysis, the metrics of TJ over its lifetime in MA is shown in Table 2. More details on the 27 metrics provided by PoP can be found in Harzing (2011).

**Table 2. Bibliometric Description of TJ during the 30-year Time Span**

| ***Metrics*** | ***Three Decades of Publication*** |
|---|---|
| Papers | 301 |
| Citations | 1,180 |
| **Coverage Years (until 2020)** | 27 |
| First year | 1993 |
| Last year | 2019 |
| Cites per Year | 43.7 |
| Cites per Paper | 3.92 |
| Cites per Author | 1,022.15 |
| Papers per Author | 268.57 |
| Authors per Paper | 1.29 |
| Cites per Author per Year | 37.85 |
| **Individual h-index** | |
| h-index | 10 |
| g-index | 30 |
| hc-index | 7 |
| hl-index | 7.14 |
| hl-norm | 10 |
| **Age-Weighted** | |
| Citation Rate (AWCR) | 164.32 |
| -index | 12.82 |
| Citation Rate per Author (AWCRpA) | 138.97 |
| e-index | 25.96 |
| hm-index | 9 |
| hl-annual | 0.37 |
| h-coverage | 65.6 |
| g-coverage | 76.9 |
| star_count | 5 |
| Estimated true Citation Count (ECC) | 1,180 |

| *Metrics* | *Three Decades of Publication* |
|---|---|
| **No. of Papers with Annual Citation Count per Year** | |
| 1 citation | 24 |
| 2 citation | 7 |
| 5 citation | 7 |
| 20 citation | 1 |

Regarding the second question, the productivity and impacts of TJ have risen over the three decades (Table 3). Only eight out of the 27 bibliometric indicators have fluctuated during the journal's 30-year history, as shown in the table.

**Table 3. Bibliometric Description of TJ per Decade**

| **Indicator** | **Publication Decade** | | |
|---|---|---|---|
| | **1990-1999** | **2000-2009** | **2010-2019** |
| Papers | 10 | 85 | 206 |
| Citations | 15 | 531 | 634 |
| Years (until 2020) | 27 | 20 | 10 |
| First year | 1993 | 2000 | 2010 |
| Last year | 1998 | 2009 | 2019 |
| Cites per Year | 0.56 | 26.55 | 63.4 |
| Cites per Paper | 1.5 | 6.25 | **3.08** |
| Cites per Author | 13 | 479 | 530.15 |
| Papers per Author | 8 | 79.81 | 180.77 |
| Authors per Paper | 1.4 | **1.19** | 1.32 |
| Cites per Author per Year | 0.48 | 23.95 | 53.01 |
| **Individual h-index** | | | |
| h-index | 2 | 7 | 7 |
| g-index | 3 | 22 | **21** |
| hc-index | 1 | 3 | 5 |
| hI-index | 2 | 6.13 | **3.77** |
| hI-norm | 2 | 7 | 7 |
| **Age-Weighted** | | | |
| Citation Rate (AWCR) | 0.63 | 38.02 | 125.66 |
| AW-index | 0.8 | 6.17 | 11.21 |
| Citation Rate per Author (AWCRpA) | 0.55 | 34.3 | 104.12 |
| e-index* | 2.24 | 19.6 | **18.03** |
| hm-index | 2 | 6.5 | **6.33** |

| Indicator | Publication Decade | | |
|---|---|---|---|
| | 1990-1999 | 2000-2009 | 2010-2019 |
| hI-annual | 0.07 | 0.35 | 0.7 |
| h-coverage | 60 | 81.5 | **59** |
| g-coverage | 73.3 | 93.2 | **71.9** |
| star_count | 0 | 1 | 4 |
| Estimated true Citation Count (ECC) | 15 | 531 | 634 |
| **No. of Papers with Annual Citation Count per Year** | | | |
| 1 citation | 0 | 3 | 21 |
| 2 citation | 0 | 3 | 4 |
| 5 citation | 0 | 3 | 4 |
| 20 citation | 0 | 0 | 1 |

*Note.* Bold numbers = fluctuating

### *Network Analysis*

The answer to the first question of the second cluster, i.e. the top ten prolific contributors of TJ and their link strength is shown in Table 4. As retrieved from MA, over the three decades there have been 282 authors contributing to TJ.

**Table 4. The Most Productive Authors in TJ during the 30-year Time Span**

| Author | Documents | Total Link Strength |
|---|---|---|
| Widiati, U. | 6 | 5 |
| Cahyono, B. Y. | 5 | 3 |
| Madya, S. | 5 | 2 |
| Djiwandono, P. I. | 5 | 0 |
| Basthomi, Y. | 5 | 0 |
| Suharmanto | 4 | 3 |
| Musthafa, B. | 4 | 2 |
| Floris, F. D. | 4 | 1 |
| Kadarisman, A. E. | 4 | 1 |
| Mukminatien, N. | 4 | 1 |

Figure 3 shows that the largest set of connected items consists of 14 items. Three of the top ten authors built up a scientific network in TJ during the investigated period.

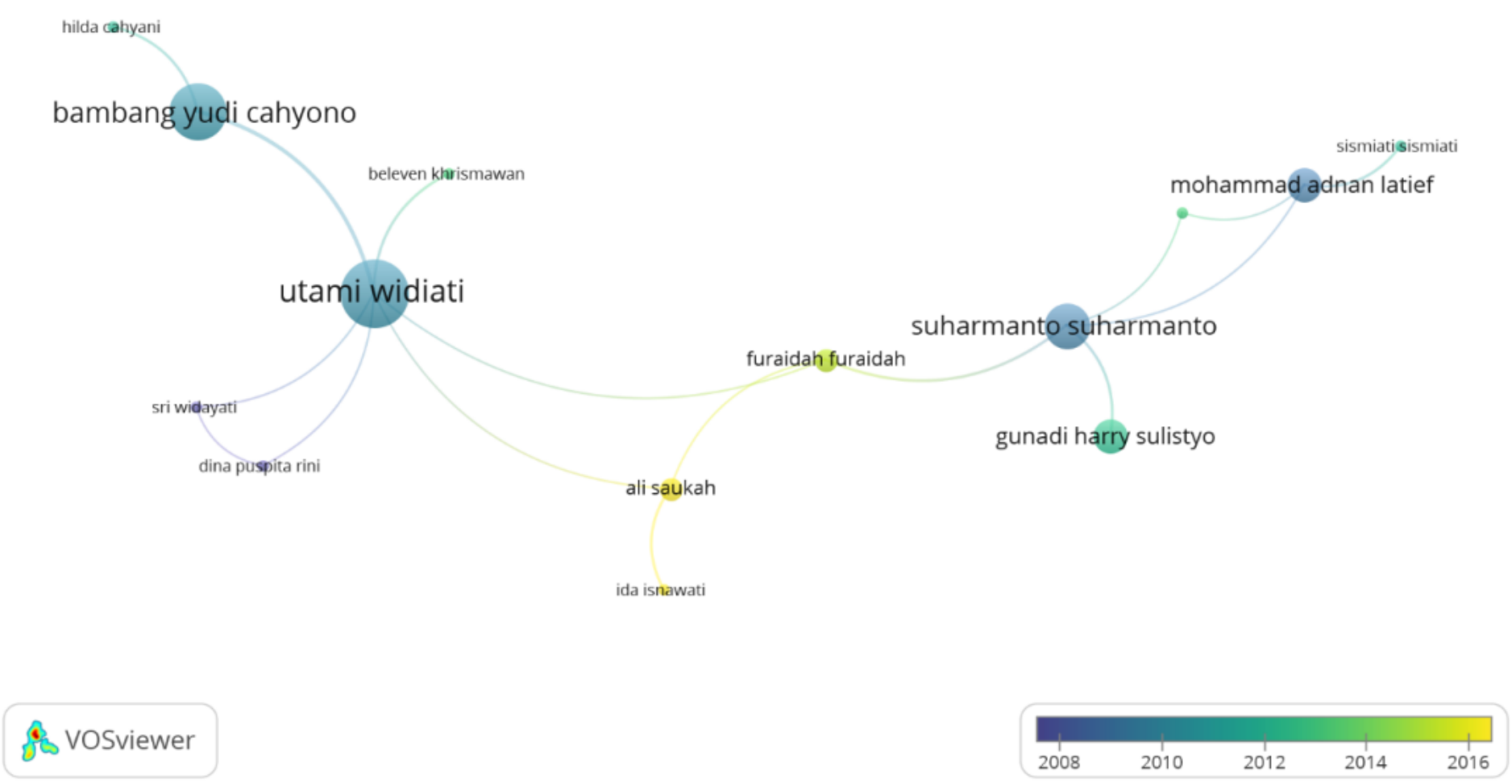

**Figure 3. The Largest Set of Co-authorship with Authors as the Unit of Analysis (Overlay Visualization)**

Table 5 shows the top ten authors arranged on the basis of productivity and co-authorship in each decade. No authors but Widiati, U. and Cahyono, B. Y. could stay in the list of top ten authors for two decades. In the third decade, having the same record, Widiati had a stronger link than Cahyono did. That is why Widiati ranked higher than Cahyono did. Both of them demonstrated their active contributions to TJ across the last two decades but they had no record in the first decade. Even from Table 5 only, it can be recognized that TJ was relatively diverse in terms of authorship.

**Table 5. The Most Productive Authors in TJ per Decade**

| 1989-1999 | | 2000-2009 | | 2010-2019 | |
|---|---|---|---|---|---|
| **Author** | **Record** | **Author** | **Record** | **Author** | **Record** |
| Musthafa, B. | 3 | Widiati, U. | 3 | Kadarisman, A. E. | 4 |
| Kam, H. W. | 1 | Madya, S. | 3 | Floris, F. D. | 4 |
| Keng, L. T. | 1 | Soedjatmiko, W. | 2 | Mukminatien, N. | 4 |
| Huda, N. | 1 | Cahyono, B. Y. | 2 | Widiati, U. | 3 |
| Sundayana, W. | 1 | Suharmanto | 2 | Cahyono, B. Y. | 3 |

| 1989-1999 | | 2000-2009 | | 2010-2019 | |
|---|---|---|---|---|---|
| **Author** | **Record** | **Author** | **Record** | **Author** | **Record** |
| Sukyadi, D. | 1 | Sutarsyah, C. | 2 | Anugerahwati, M. | 3 |
| Hall, E. | 1 | Algadrie, L. | 2 | Alberth | 3 |
| Lestari, L. A. | 1 | Djiwandono, P. I. | 2 | Mistar, J. | 3 |
| Amarien, N. | 1 | Harsono, Y. M. | 2 | Astika, G. | 3 |
| Wardoyo, S. | 1 | Basthomi, Y. | 2 | Mambu, J. E. | 3 |

Regarding the third question, there were 42 institutions from 9 countries contributing to TJ over the study period. As can be seen in Figure 4, with its 32 articles, Universitas Negeri Malang (UM), was the most active contributing institution. The maximum number of papers by other institutions was five papers by the Catholic University of America (USA) followed by Satya Wacana Christian University (Indonesia) producing 4 articles.

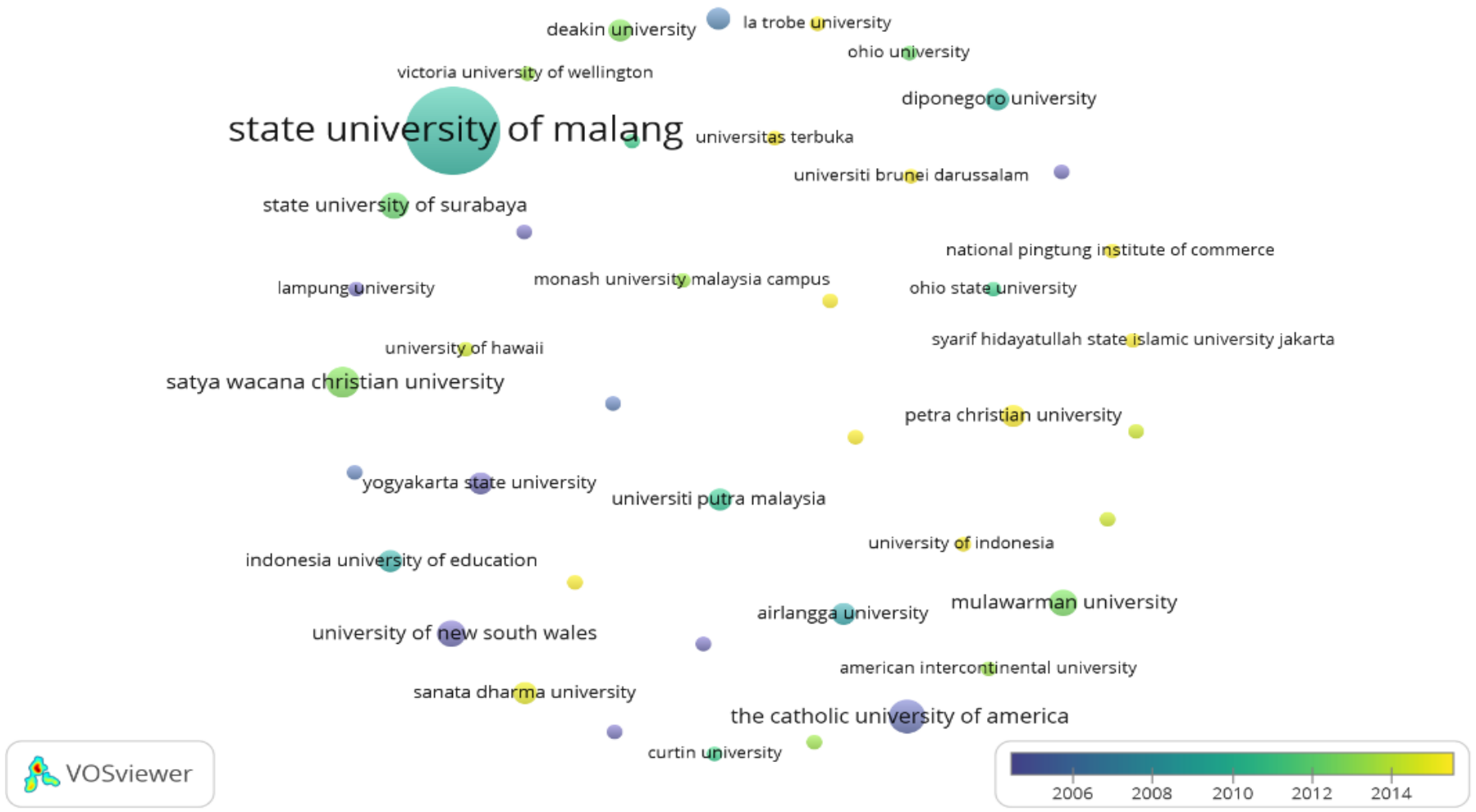

**Figure 4. Co-authorship Map based on Organizations (Overlay Visualization)**

As the fourth question concerns the most productive countries in TJ, having 18 institutions as the contributors to TJ, Indonesia was the most productive countries in TJ. The nine countries are listed in Table 6.

Interestingly, within 15 contributing nations, the countries where English is used as neither a foreign nor second language such as the United States of America constitute nearly 36% of the contributing ones. It also shows the diversity of contributing countries in TJ.

**Table 6. The Most Productive Countries in TJ**

| Country | Contributing institution | % |
|---|---|---|
| Indonesia | 18 | 42.86 |
| Australia* | 6 | 14.29 |
| USA* | 6 | 14.29 |
| Malaysia | 4 | 9.52 |
| New Zealand* | 3 | 7.14 |
| Iran | 2 | 4.76 |
| Brunei Darussalam | 1 | 2.38 |
| Singapore | 1 | 2.38 |
| Taiwan | 1 | 2.38 |

**Note*: Countries where English is commonly used as a first language.

After the co-authorship analysis with authors and organization/ countries as the units of analysis, the fifth question was about the pattern of authorship in TJ. Table 7 shows the number of articles based on the number of authors. Interestingly, in 2004 one article entitled "Developing a model of teaching English to primary school students". was coauthored by seven authors including Madya. In 2019 Madya also coauthored one of the three articles by five coauthors, i.e. "The equivalence of TOEP forms". During the journal's lifetime, over 80% of articles were authored by a single author.

**Table 7. The Pattern of Authorship in TJ**

| No of author(s) | Frequency | Percent |
|---|---|---|
| 7 | 1 | 0.33 |
| 5 | 3 | 1.00 |
| 4 | 2 | 0.66 |
| 3 | 10 | 3.32 |
| 2 | 41 | 13.62 |
| 1 | 244 | 81.06 |

To answer the sixth question related to the top-cited articles and the article types in TJ, the titles and abstracts were mined from MA. Table 8 indicates the top ten cited articles in TJ over the three decades. No record could be found from the first decade. Eight out of the top cited articles were conceptual articles. Fifty-three percent of the total citations in Table 8 were from those published in the second decade.

**Table 8. The Top Ten Cited Articles in TJ and the Article Types**

| Authors | Title | Year | Cites | Annual Cites | Type |
|---|---|---|---|---|---|
| Lie, A. | "Education policy and EFL curriculum in Indonesia: between the commitment to competence and the quest for higher test scores" | 2007 | 196 | 15.08 | Conceptual |
| Marcellino, M. | "English Language Teaching in Indonesia: a continuous challenge in education and cultural diversity" | 2008 | 139 | 27.80 | Conceptual |
| Cahyono, B. Y. and Widiati, U. | "The teaching of EFL listening in the Indonesian context: the state of the art" | 2009 | 96 | 6.86 | Conceptual |
| Nord, C. | "Translating as a purposeful activity: A prospective approach" | 2006 | 94 | 6.71 | Conceptual |
| Hamra, A. and Syatriana, E. | "Developing a model of teaching reading comprehension for EFL students" | 2010 | 76 | 15.20 | Empirical |
| Harsono, Y. M. | "Developing learning materials for specific purposes" | 2007 | 66 | 13.20 | Conceptual |
| Cahyono, B. Y. and Widiati, U. | "The teaching of EFL vocabulary in the Indonesian context: The state of the art" | 2008 | 66 | 13.20 | Conceptual |

| Authors | Title | Year | Cites | Annual Cites | Type |
|---|---|---|---|---|---|
| Megawati, F. and Anugerahwati, M. | "Comic strips: A study on the teaching of writing narrative texts to Indonesian EFL students" | 2012 | 11 | 1.38 | Empirical |
| Campbell, S. | "Translation in the context of EFL - the fifth macroskill?" | 2002 | 10 | 0.56 | Conceptual |
| Madya, S. | "Developing standards for EFL in Indonesia as part of the EFL teaching reform" | 2002 | 10 | 0.56 | Conceptual |

The titles and abstracts retrieved from MA were also used to answer the seventh question of the network analysis. In this case, the unit of analysis was the fields of study. The top ten keywords occurred in the 301 titles and abstracts were psychology with 282 occurrences, pedagogy (219), linguistics (143), Indonesian (69), teaching method (53), curriculum (33), English as a second language (25), foreign language (25), language acquisition (24), and grammar (22). Table 9 compares the top ten keywords extracted from the 301 titles and abstracts per publishing decade of TJ. The top three keywords in the three decades were also the top three ones in each of the decades.

**Table 9. The Top-used Keywords in TJ per Decade**

| 1989-1999 (10 documents, 34 keywords) | | 2000-2009 (85 documents, 299 keywords) | | 2010-2019 (206 documents, 633 keywords) | |
|---|---|---|---|---|---|
| **Keyword** | **Freq.** | **Keyword** | **Freq.** | **Keyword** | **Freq.** |
| psychology | 10 | psychology | 80 | psychology | 192 |
| linguistics | 8 | pedagogy | 53 | pedagogy | 160 |
| pedagogy | 6 | linguistics | 47 | linguistics | 88 |
| basic writing | 1 | foreign language | 12 | teaching method | 47 |
| comprehension approach | 1 | grammar | 10 | curriculum | 24 |
| computerized adaptive testing | 1 | curriculum | 9 | English as a second language | 24 |

| 1989-1999 (10 documents, 34 keywords) | | 2000-2009 (85 documents, 299 keywords) | | 2010-2019 (206 documents, 633 keywords) | |
|---|---|---|---|---|---|
| content area | 1 | vocabulary | 9 | language acquisition | 20 |
| critical reading | 1 | first language | 7 | language proficiency | 19 |
| early reading | 1 | active listening | 6 | second language instruction | 18 |
| engineering ethics | 1 | sociology | 5 | language education | 15 |

The last two questions of the second cluster concern the relatedness of articles in TJ viewed from the frequency they cited at each other and from the number of shared references. The results of citation analysis with documents as the unit of analysis is shown in Figure 5. It indicates a relatively few number of citation links between the articles. In other words, the majority of articles in TJ did not cite articles published earlier in TJ.

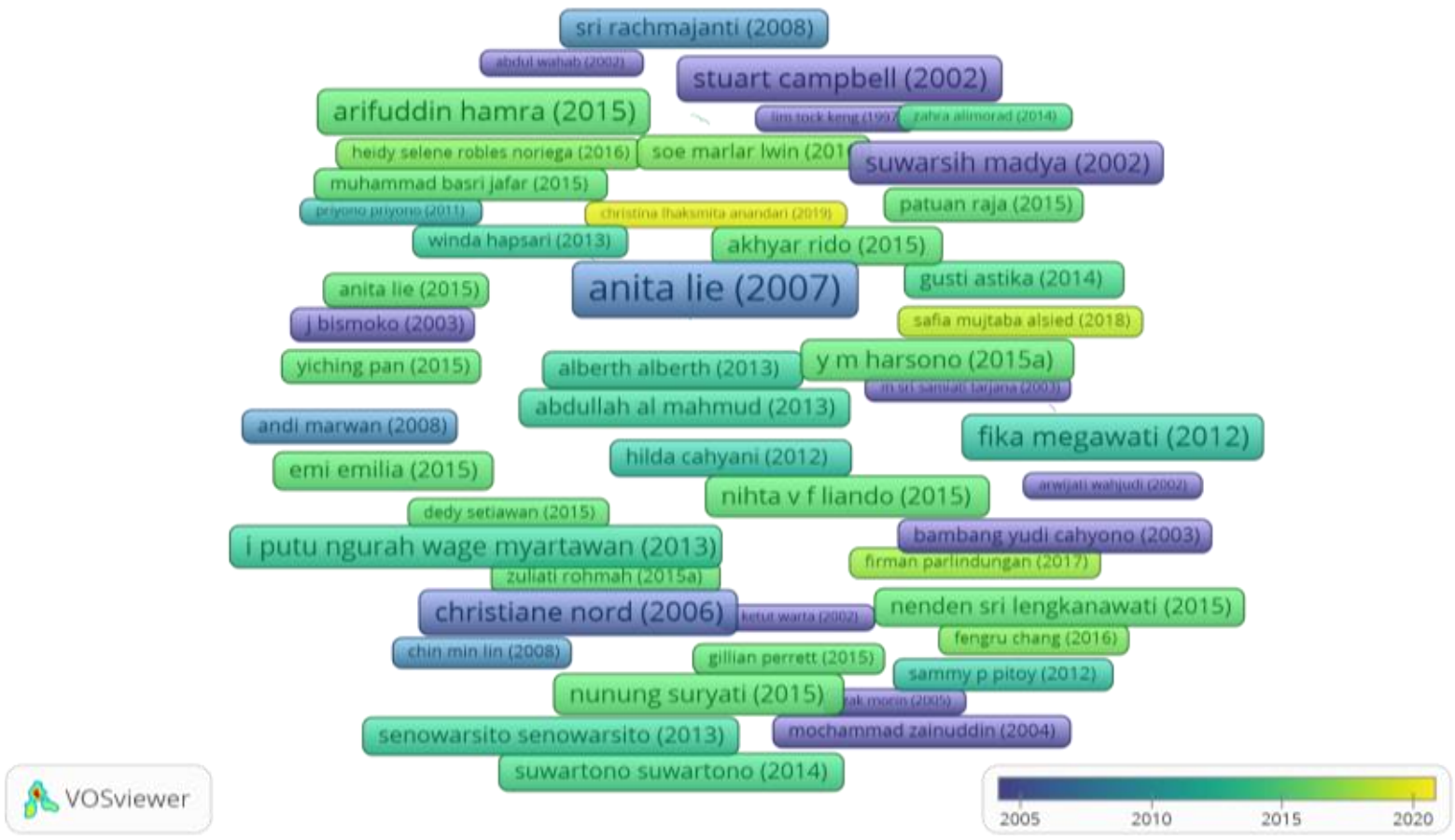

**Figure 5. Relatedness of Documents based on the Number of Co-citation (Overlay Visualization)**

A more careful inspection reveals the largest set of only 13 connected documents as automatically ordered by VOSviewer in terms of co-citation within TJ (Table 10). Only four documents in the five clusters of journal's self-citation were written by the same authors. At an author level, no author but Musthafa, B. did a self-citation. In his article published in 2011 (or 2015 according to MA), he cited his own article that had also been published in TJ some years earlier. i.e. in 1996 (or 1997 in MA).

**Table 10. The Relatedness of Articles in TJ based on the Citation Frequency**

| Authors(s) | Year | Title |
|---|---|---|
| *Cluster 1* | | |
| Cahyono, B. Y., and Widiati, U. | 2009 | "The teaching of EFL listening in the Indonesian context: the state of the art" |
| Lie, A. | 2007 | "Education policy and EFL curriculum in Indonesia: Between the commitment to competence and the quest for higher test scores" |
| **Astuti, S. P.** | 2013 | "Teachers' and students' perceptions of motivational teaching strategies in an Indonesian high school context" |
| *Cluster 2* | | |
| Collins, P. | 2006 | "Grammar in TEFL: A critique of Indonesian high school textbooks" |
| Madya, S. | 2007 | "Searching for an appropriate EFL curriculum design for the Indonesian pluralistic society" |
| Lie, A. | 2017 | "English and identity in multicultural contexts: Issues, challenges, and opportunities" |
| *Cluster 3* | | |
| **Musthafa, B.** | 1997 | "Content area reading: Principles and strategies to promote independent learning" |
| **Musthafa, B.** | 2001 | "Communicative Language Teaching in Indonesia: Issues of theoretical assumptions and challenges in the classroom practice" |
| **Astuti, S. P.**, and Lammers, J. C. | 2017 | "Making EFL instruction more CLT-oriented through individual accountability in cooperative learning" |
| *Cluster 4* | | |
| Masduqi, H. | 2011 | "Critical thinking skills and meaning in English language teaching" |
| Marcellino, M. | 2008 | "English Language Teaching in Indonesia: a continuous challenge in education and cultural diversity" |

| *Cluster 5* | | |
|---|---|---|
| Larson, K. R. | 2014 | "Critical pedagogy(ies) for ELT in Indonesia" |
| Hayati, N. | 2010 | "Empowering non-native English speaking teachers through critical pedagogy" |

After the analyses of co-authorship, citation co-occurrence, and co-citation, it is also important to look at the shared references the authors cited. With the fractional counting method and documents as the unit of analysis, the analysis of bibliographic coupling shows that only 79 out of the 301 articles in TJ were related in terms of the references coupled in the articles (Figure 6). The links between two authors demonstrate the shared references or bibliographic coupling in their articles.

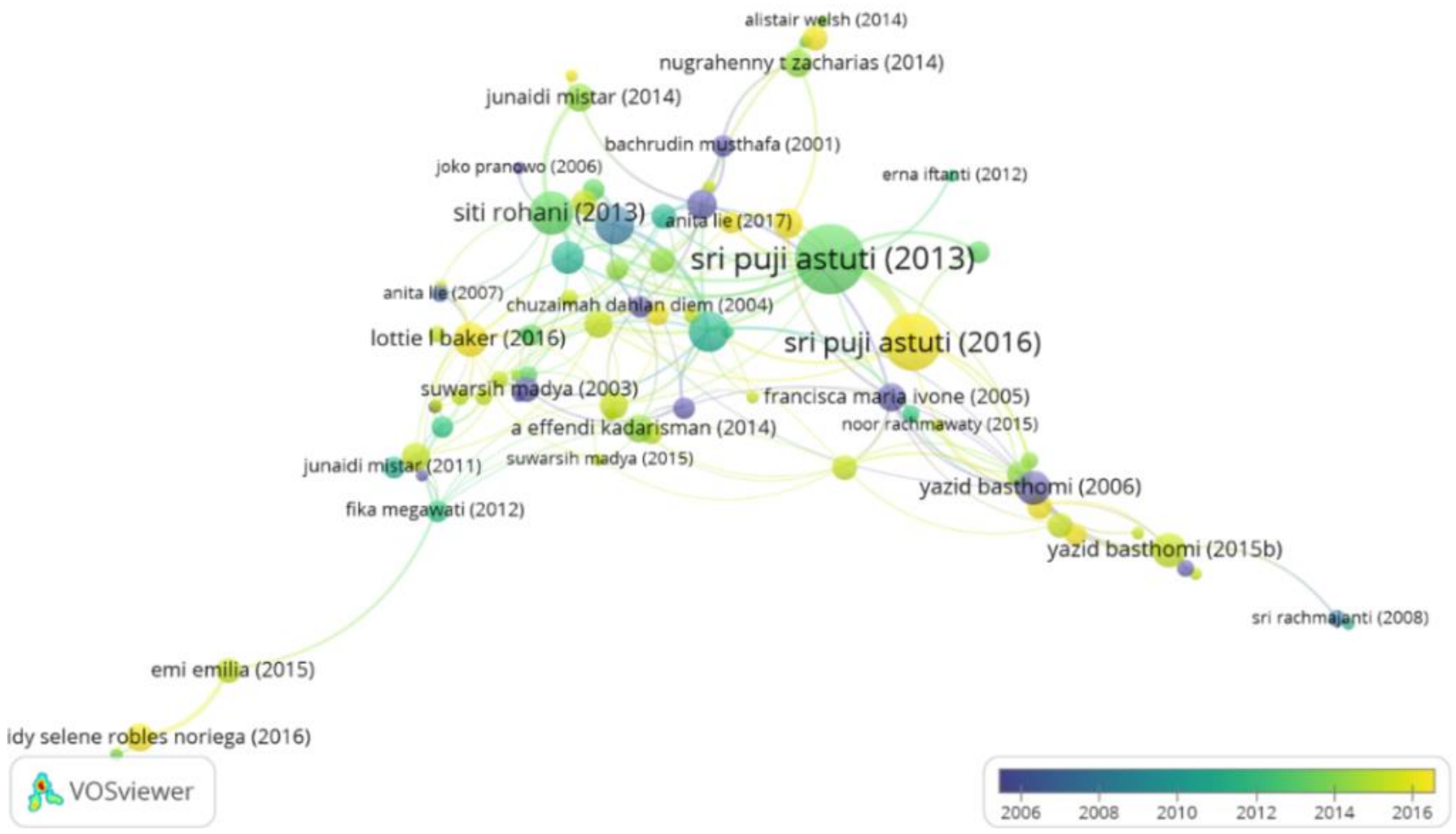

**Figure 6. Relatedness of Articles in TJ viewed from the Number of Shared References (Overlay Visualization)**

The top ten articles in terms of bibliographic coupling can be seen in Table 11. The articles were listed in order of the number of references cited in two articles. Five out of the ten articles having the highest number of same references in TJ were (co)authored by two authors, i.e. Astuti, S.P. and Basthomi, Y. Interestingly, the two articles by Basthomi, Y. were published in a relatively longer time span than those by Astuti, S.P.

**Table 11. The Relatedness of Articles in TJ based on the Number of Shared References**

| Author(s) | Year | Title | Strength |
|---|---|---|---|
| Astuti, S. P. | 2013 | "Teachers' and students' perceptions of motivational teaching strategies in an Indonesian high school context" | 33 |
| Astuti, S. P. | 2016 | "Exploring motivational strategies of successful teachers" | 22 |
| Rohani, S. | 2013 | "Positive versus negative communication strategies in task-based learning" | 13 |
| Abdullah, U. | 2011 | "EIL in practice: Indonesian and Chinese international postgraduate students negotiate meaning" | 13 |
| Tulung, G. J. | 2013 | "Oral discourse generated through peer-interaction while completing communicative tasks in an EFL classroom" | 11 |
| Baker, L. L. | 2016 | "Re-conceptualizing EFL professional development: Enhancing communicative language pedagogy for Thai teachers" | 8 |
| Basthomi, Y. | 2006 | "Rhetorical odyssey and trajectories: A personal reflection" | 8 |
| Basthomi, Y. | 2015 | "Examining research spaces in doctoral prospectuses" | 8 |
| Baleghizadeh, S., and Oladrostam, E. | 2011 | "Teaching grammar for active use: A framework for comparison of three instructional techniques" | 7 |
| Astuti, S. P., and Lammers, J. C. | 2017 | "Making EFL instruction more CLT-oriented through individual accountability in cooperative learning" | 6 |

## Discussion

### *Article Record, Indexation and Citation*

Through the lens of MA, this study captures the bibliometric portrait of TJ during its 30-year time span. Unexpectedly, as displayed in the Findings subsection, the journal's website shows the earliest issue was the Volume 10 Number 1 in 1999 whereas ERIC indicates the earliest publication year was in 1993 (Table 3). Based on the average article of seven per issue, a rough estimate of missing articles between 1990 and 2000 put the number of articles at 145.

Fortunately, two of the missing articles could be found in ERIC. They could be verified and found valid. One of them is "Teaching EFL learners sociolinguistic concepts for intercultural understanding" coauthored by Musthafa, B and Sundayana, W in September 1993 (ERIC, 2020). In this case, both of the authors took the initiative to submit the article to and granted ERIC the permission to digitize and share it. The fact that there were some missing articles could explain why there were only ten articles recorded in the first decade (Table 3).

Another possible explanation for the relatively less than expected number of articles in the first decade is that some articles published in the first decade were indexed in the third decade. This also happened to some articles published in the second decade.

In fact, as shown in its website, TJ has put a lot of effort into digitizing the pre-electronic articles. Because the scanning and journal publishing technologies such as OJS are affordable nowadays, the incomplete digitizing project going back to the first volumes could possibly be explained by the loss of the pre-electronic articles due to a relatively inappropriate archiving process. That the articles can no longer be read and cited is feared by not only authors but also editors and publishers (Morris et al., 2013). A lot of older research studies, especially those published in TJ as one of the pioneering journals in the field of ELT in Indonesia and Southeast Asia, possibly remain relevant and important for the contemporary advancement of the field.

Two viable solutions for the loss was offered by Morris et al., (2013). Firstly, TJ should contact the authors and their affiliations or some libraries, societies, and individuals to find, donate, or loan the related work in a print or electronic form. Another solution is by posting a notice about the digitization project of back issues on the journal's website and social media. The solutions, at least, could help minimize the permanent loss of the scholarly work. The next few years could see a more complete set of TJ.

In terms of indexation, TJ is a great success. The journal's apparent success, however, should be improved by updating the journal's records in some indexing services. Some indexing services such as Google Scholar and MA use special software to update their databases automatically based on the latest data fed by a publisher, whereas some other indexing ones such as DOAJ and Scopus need a journal's active role in curating its own metadata (Gusenbauer & Haddaway, 2020; Hug et al., 2017; Thelwall, 2017). From a closer inspection of Table 1, it emerges that the latest issue searchable in DOAJ is the Volume 27 Issue 1. More precisely, at this study period, the last time the record of TJ updated was in February 2017 (DOAJ, 2020).[2] As one of the pillars of scientific performance at all levels (Nylander et al., 2020), the indexation often means visibility, readership, and reputation (Marotti de Mello & de Sandes-Guimarães, 2019). Paying more attention to important aspects of indexation such as providing the correct metadata and Digital Object Identifiers could lead to higher visibility, a broader readership, and better reputation.

The last three decades have witnessed the journal's commitment and competence in the scholarly publication. In terms of 19 out of 27 metrics in PoP software (Harzing, 2011), the journal's publication and citations have risen during its lifetime. The eight fluctuating metrics in the three decades are Cites per Paper (CpP), Authors per Paper, g-index, hI-index, e-index, hm-index, h-coverage, and g-coverage.

However, as the metrics were complicated by some inaccurate metadata and indexations, the fluctuating metrics should be interpreted with considerable caution. For example, the first fluctuating bibliometric indicator, CpP, is the number of total citations divided by the number of total papers in a journal for a certain period (Harzing, 2011). Usually, CpP grows through the time (Merigó et al., 2018). The earlier published articles generally receive more citations. The inappropriate metadata leading to misplaced indexation could be one of the reasons for all of the fluctuating bibliometric indicators.

Estimated on the basis of the last two publication decades (Table 3), TJ published around 15 articles per year. Unlike TJ that has shown the "consistency" of 15 articles per year (7 or 8 articles per publication number), some top journals in the relevant fields such as System (Lei & Liu, 2019b) have shown the "inconsistency" of the annual number of articles. In 2017, for example, System published 80 articles but in the following two years the publication increased by 20% (System, 2020). The "consistency" was possibly as a result of financial challenges (Silver, 2018) such as no submission fee and article processing charges (TEFLIN Journal, 2020) and relatively limited financial aids from the government (Wiryawan, 2014). Even though information about the rejection rate was hardly found, the seven or eight articles per publication could also mean the high quality of rigorous editorial process.

---

[2] Based on e-mail correspondence with the editors of TEFLIN Journal during the preparation for this article publication, the TEFLIN Journal record in DOAJ has now been updated until the most recent publication.

Being selective in publishing articles results in a few articles which still possibly receive a very high rate of CpP. The low quantity or publication, however, could lead to the low quality or impact. As shown in Table 3, CpP as one of the single-number metrics changed drastically, from 6.25 to 3.08, a decrease of -200% but the h-index and g-index as two of the multi-number metrics combining publications and citations did not, i.e. around seven and 21, respectively. Providing an example of a journal reaching a maximum h-index of 50 after one decade for publishing only five articles a year, Harzing (2011) confirms that the limited publication will limit a journal's overall impact. Achieving a good balance between the quantity (publication) and quality (citation), needless to say, is important to increase a journal's impact in a certain field.

This work portrays the bibliometric portrait of publication, indexation, and citation of TJ as one of the leading journals in Indonesia. While a great body of bibliometric analyses at the journal level found a complete set of correctly indexed publications to analyze (e.g. Martínez-López et al., 2018), this study faced a challenge of some missing articles from the first decade of the journal publication and incomplete indexing. This reveals one of the areas to improve by TJ's gatekeepers in order to show their commitments to the enterprise of ELT and, especially, to the authors' academic contributions.

A replication study could be conducted in another milestone such as the golden jubilee or after TJ is able to publish and index all of the back articles. The replicability of this study, one of the most important aspects in the bibliometric research (Gusenbauer & Haddaway, 2020), was ensured as the portrait was captured by the PoP (Harzing, 2007), a free and user friendly bibliometric application, through the lens of MA, one of the free ASEBDs widely used in the bibliometric arena (Gusenbauer & Haddaway, 2020; Harzing, 2019; Hug et al., 2017; Thelwall, 2017). Other previous studies mining bibliometric data from Scopus, e.g. Barrot (2017), and WoS, e.g. Zhang (2020), possibly have lower transparency and replicability as the two databases could be accessed with a subscription only. MA also covers all of the publication years and citations, not limited in the coverage years and citations by other journals in the paid databases.

The three decades has therefore witnessed the journal's publication, indexation, and citation. The examined single- and multi-number metrics of bibliometric quantity and quality could enable readers, prospective authors, and the journal's gatekeepers see them in accordance with their needs. Seen in the bibliometric metrics including the publication year of 30, it is legitimate to say that TJ has relatively succeeded in "enhancing and disseminating scholarly work in the field of ELT" (TEFLIN Journal, 2020).

***Network of Co-authorship, Citation, Co-occurrence, and Bibliographic Coupling***

To exhibit a more vivid bibliometric portrait of TJ, this study adopted the *VOSviewer* (Eck & Waltman, 2020) to catch the connectedness of publication in terms of co-authorship, citation, co-occurrence, bibliographic coupling, and co-citation through the lens of MA. The use of same database to answer the second cluster of questions could ensure the consistency of analysis. As previously discussed, the free abstract and citation database along with the free highly innovative software make this study comprehensive in terms of the coverage, replicable and transparent in terms of the data retrieval (Gusenbauer & Haddaway, 2020). The three requirements could hardly be met by the previous studies because some subscription-based databases afford limited researchers whose access is institutionally provided.

The first five questions of the second cluster were concerned with the co-authorship. As Merigó et al. (2018) describe, the co-authorship identifies the network of authors and their affiliations or countries. The descriptive analysis (Authors per Paper in Table 3) and network analysis (Table 7) reveals the dominance of sole authorship in TJ. Such a dominance in the field of (applied) linguistics has been reported by Barrot (2017). Harzing (2011) also found the same pattern in the two subject areas falling within the scope of TJ in Scopus (2020).

Despite the dominance of sole authorship, in terms of productivity, only two out of the top ten productive authors co-authored no papers. Of the 282 authors, Widiati and Cahyono had not only the most papers but also the strongest co-authorship links. To conclude that the co-authorship is positively related to productivity, however, certainly needs further investigations.

In relation to the organization as the analysis unit, UM was the most active contributing institution. However, papers from authors in the institution have been published apparently without bias as only five of the 35 reviewers and three of the seven editors are from the university (TEFLIN Journal, 2020). Another evidence is that the 32 records from the university's scholars during the three decades constitutes only about 10% of the total publication. The relatively low contribution could be explained by the fact the scholars from the university published more papers in other (more internationally reputable) academic venues as found by Barrot (2017).

The data retrieved from MA show that the most productive country during the 30 years of TJ is Indonesia. Two possible reasons for this finding were noted by Barrot (2017). Firstly, some top regional and international journals have the higher rejection rate compared to TJ. Because of being one of the most enduring journals in Indonesia and Southeast Asia along with its annual publication of 15 or 16 articles, TJ clearly has a relatively low acceptance rate. A more acceptable explanation is that TJ is one of the most appropriate scholarly venues for voicing concern about the teaching of English as a foreign language in Indonesian context. During its lifetime TJ has witnessed how since the Indonesian independence the challenging educational and socio-cultural diversity in the teaching of English as a foreign language in Indonesia has been complicated by, at least, unsuccessful four curriculum changes (Lie, 2017; Marcellino, 2008). The changes the policy makers imposed are sometimes not based on by theoretically driven and research-based considerations (Lie, 2007).

In TJ, the articles voicing concern about ELT in the EFL context are classified as conceptual articles. This type of scholarly writing loosely defined as "just papers without data" includes theoretical, review, commentary, and critique writing (Gilson & Goldberg, 2015, p. 127). Interestingly, eight of the top ten cited articles in TJ belong to the non-empirical ones. The similar scenario was also observed by Lei and Liu (2019a, 2019b). However, a closer reading of the conceptual work in Table 8 reveals that seven out of them voiced concern about ELT in Indonesia. That the views or critiques are everlasting could explain why the non-empirical type has attracted more widespread interest than its counterpart has in TJ.

The pattern of citation in TJ is also interesting to discuss. As can be seen in the online supplementary materials (Syahid and Mukminatien, 2020), the total publication of TJ from 1990 to 2019 that could be indexed in MA is 301 papers of which 174 papers received at least one citation. Twenty percent of the total cited papers, i.e. 60 of 174 papers, accounts for 86.13% of total citations, i.e. 1,006 of 1,168 citations. To put it simply, over 80% of the total citations came from 20% of the cited papers in TJ. The pattern of 80/20 follows the Pareto distribution, also known as the power law or Zipf's law, widely applied in many fields and contexts from planetary to social sciences (Newman, 2005). To put it simply, 80% of the effects results from 20% of the causes.

Related to the analysis of co-occurrence, since the first decade TJ has witnessed steady keywords. TJ seems to revolve around psychology, pedagogy, and linguistics as identified from the 301 titles of which 287 accompanied with their abstracts. Two decades before TJ was launched, the three variables as the "trinity or unity" in the field of (English as a foreign or second) language teaching had been excellently argued by Wardhaugh (1968, p. 80). For over the four decades most of the top discussed topics in System such as identity and self-efficacy (2019b) have been under the umbrella of the trinity. Within this view, TJ has voiced not only the local and regional concern but also the global one.

Regarding the co-citation analysis with documents as the unit of analysis, only 13 out of 301 (4.33%) articles in TJ receiving one or more citations from other articles in TJ. In other words, the rate of self-citation in which the citing and cited articles published in TJ is very low. Such a low level is very encouraging since the high level of self-citation in a journal often arises suspicions and might lead to a journal's exclusion from such dominant players in indexing services as WoS (Oransky, 2020). The practice of self-citation at the journal level that could help increase a journal's bibliometric performance is "very common for *the majority of journals*" (Merigó et al., 2018, p. 258, emphasis added). However, as Mahian and Wongwises (2015) caught, editors of some journals published by well-known academic publishers made citing papers previously published in their journals optional, semi-mandatory, or even mandatory for those submitting manuscripts. The low self-citation in TJ could possibly be explained by the fact that there is no way the gatekeepers of TJ do the relatively debatable unethical practices.

The last analysis is concerned with bibliometric coupling. In contrast to the topical uniformity of psychology, pedagogy, linguistics, and pedagogy in the key-word co-occurrence, the observed same references cited by two articles published in TJ were relatively more diverse. About 75% of the documents cited by the authors in TJ was different from each other. The results thus emphasize how most of the authors have given the readers of TJ different perspectives on the above trinity.

## CONCLUSIONS

In 2019 TJ celebrated the 30th anniversary. To examine whether TJ has achieved what it originally envisaged in the arena of ELT, a study was conducted involving descriptive and network analyses of bibliometric data from MA, a free abstract and citation database, using two pieces of free software, PoP and VOSviewer. The three decades have witnessed how TJ grows around the trinity of publication, indexation, and citation. The unity of commitment and competence of the journal's gatekeepers including the authors have resulted in the national, regional, and global recognition.

TJ can achieve higher impact by maintaining such good managerial and editorial standards including the consistent focus on the psychology, pedagogy, and linguistics as the pillars of ELT from diverse perspectives by the authors of diverse institutional and national backgrounds with zero tolerance of unethical practices. It could also be done by providing full sets of back issues through collaboration with some institutions, libraries, authors, and readers. Paying closer attention to the indexing process could help the journal reach higher visibility. Another thing to underline is the complexity of relatively unavoidable multi-number metrics in which not only the quality of articles (citations) but also the quantity of them should be well balanced. In other words, more articles published by TJ which necessarily means more financial supports and heavier editorial workloads could increase the journal's impact, especially to overcome the fierce competition between academic journals for authors and readers' attentions.

This study captures the bibliometric character of TJ as one of not only the pioneering Open Access Journals but also leading journals in the field both nationally and regionally. The portrait exhibited is basically spatial temporal as the data was derived from MA only within the aforementioned limited timespan. Despite this, the use of same database for the two different pieces of free software could ensure the coverage, consistency, replicability, and transparency of this study. The dynamics of bibliometric data of TJ could serve as a stimulus for future research, for instance by adopting other databases and computer programs along with examining the relationship between research metrics. Such studies including this work offer not only retrospective but also prospective insights into the scientific development of ELT and the journal's future direction.

.